\documentclass[a4paper,twocolumn,11pt,accepted=2025-02-19]{quantumarticle}
\pdfoutput=1

\expandafter\let\csname equation*\endcsname\relax
\expandafter\let\csname endequation*\endcsname\relax

\usepackage{enumerate,appendix}
\usepackage{amsmath, amsthm, amssymb,commath}
\usepackage{color,calc,graphicx}
\usepackage[usenames,dvipsnames,svgnames,table,cmyk,hyperref]{xcolor}
\usepackage[colorlinks]{hyperref}
\if0
\hypersetup{
	colorlinks = true,
	urlcolor = {blue},
	citecolor = {blue},
	linkcolor= {blue}
}
\fi

\usepackage{graphicx}
\usepackage{amsmath}
\usepackage{latexsym}
\usepackage{bbm}

\usepackage{multirow}

\newcommand{\SU}{\mathop{\rm SU}\,}

\def \be {\begin{equation}}
\def \ee {\end{equation}}

\newcommand{\Tr}{\mathrm{Tr}}

\def \sofc2{{\cal S}({\mathbb C}^2)}

\def\>{\rangle}
\def\<{\langle}

\theoremstyle{plain}

\def\Label#1{\label{#1}\ [\ \text{#1}\ ]\ }
\def\Label{\label}

\begin{document}

\title{Heisenberg scaling based on population coding}

\author{Masahito Hayashi}
\address{School of Data Science, The Chinese University of Hong Kong,
Shenzhen, Longgang District, Shenzhen, 518172, China}
\address{International Quantum Academy, Futian District, Shenzhen 518048, China}
\address{Graduate School of Mathematics, Nagoya University, Furo-cho, Chikusa-ku, Nagoya, 464-8602, Japan}
\orcid{0000-0003-3104-1000}
\email{hmasahito@cuhk.edu.cn, masahito@math.nagoya-u.ac.jp}
\maketitle

\begin{abstract}
We study Heisenberg scaling of quantum metrology
in the viewpoint of 
population coding.
Although Fisher information has been used for a figure of merit 
to characterize Heisenberg scaling in quantum metrology,
several studies pointed out it does not work as a figure of merit
because it does not reflect the global structure.
As an alternative figure of merit,  
we propose the mutual information, which connects
the number of distinguishable elements of the parameter space
in the viewpoint of population coding.
We show that several unitary models achieve Heisenberg scaling in this context.
\end{abstract}


\maketitle

\section{Introduction} %--Problem with Fisher information}
%\cite{BHBMPW},
Quantum metrology is a key technology for estimating physical parameters
in quantum systems with high resolution.
It is known that Fisher information gives
the precision limit of estimation under the $n$-copies of the unknown state is available \cite{Helstrom,Holevo,two2,two4,HM08,HO}.
Due to the above successful role of Fisher information,
many people believed that Fisher information gives
the precision limit of estimation even in the quantum metrology including the unitary estimation.
Therefore, many people consider that 
the noon state realizes Heisenberg scaling in the quantum metrology because 
it achieves the maximum Fisher information \cite{Giovannetti11,Giovannetti06,Giovannetti,Jonathan,Imai,Okamoto_2008,Nagata07,Thomas-Peter} in the the unitary estimation.
Although the variance-type of error is lower bounded by this maximum Fisher information,
this lower bound cannot be achieved in the sense of 
global estimation 
because this lower bound is strictly larger than
the minimum error 
under the Bayesian criterion or the minimax criterion 
\cite{BDM,Ha06,H11,HVK,HLY,gorecki2020pi,KVSHZ}.
This fact shows that Fisher information does not reflect the global information structure.
Although the references \cite{Ha06,H11,HVK,HLY,gorecki2020pi} pointed out this problem for Fisher information, 
still several papers \cite{YZL,LHYY,Liu2021,Fan24,Hu22,QIN2024129795,Triggiani} use Fisher information as a figure of merit for quantum metrology.
To resolve this problem, it is necessary to propose an alternative figure of merit to reflect global information structure.
%Therefore, it is needed to seek another information quantity to measure the amount of the information in the parameter estimation.
In fact, finding such an alternative figure of merit has remained as an open problem in \cite{HVK}.
As an alternative figure of merit, this paper focuses on 
the mutual information between the parameter and the $n$-fold tensor product quantum system of our interest.

Usually, the mutual information is used for channel coding problem.
In the classical case, the idea of population coding was proposed in the context of a neural field \cite{Wu,Panzeri2010}.
It presented the idea that the information is presented by
an ensemble, i.e., a population, of neurons subject to the same distribution in a neural field.
Mathematically, we consider a family of distributions 
$\{p_\theta\}_{\theta\in \Theta}$ with $\Theta \subset \mathbb{R}^d$,
and a population, i.e., an $n$ random variables $\vec{X}=(X_1,\ldots,X_n)$ are subject to the $n$-fold independent and identical distribution $p_\theta^n$ of $p_\theta$. 
The meaningful information $\theta$ is coded by using the distribution $p_\theta^n$. 
In this context, we focus on
the mutual information $I(\vec{X};\theta)$ between
$\vec{X}$ and $\theta$ under a prior distribution $\mu$ 
of the parameter $\theta$.
It is known that the logarithm of the number of distinguishable elements of 
the parameter set $\Theta$ is upper bounded by
the mutual information $I(\vec{X};\theta)$.
Hence, the mutual information $I(\vec{X};\theta)$ plays a key role in the population coding.
In fact, the references \cite{KB1,KB2} showed that  
the mutual information $I(\vec{X};\theta)$ 
behaves as $\frac{d}{2}\log n$ in the context of Bayes codes.
Unlike Fisher information, the mutual information reflects 
the global information structure.
The reference \cite{H09} studied the same problem in the quantum setting.
The references \cite{YCH,HT18,YBCH} focused on this type of mutual information 
in the context of data compression for $\theta$ in the classical and quantum setting.

In this paper, we study 
the unitary estimation with $n$-parallel application of the same unitary under several unitary models $\{U_\theta\}_{\theta \in \Theta}$. 
As the first topic, when $d$ is the number of parameters, 
we show that the mutual information behaves as $d\log n$, which is twice as the standard scaling $\frac{d}{2}\log n$.
As the second topic, to follow the concept of population coding, 
we discuss how many elements in the parameter space $\Theta$ can be distinguished under $n$-parallel application.
We show that the logarithm of the number of distinguishable elements is also $d \log n$ with the best initial state preparation.  
As discussed in Sections \ref{S2} and \ref{S33},
the number of distinguishable elements has useful meaning in estimation.
Another problem is the necessity of use of the ancilla or the reference system for the input state.
This problem has been studied in \cite{Layden,Zhou}.
We consider this problem based under our problem setting.

The preceding studies discussed 
the first topic for the phase estimation \cite{Hall,Hassani} and multi-phase estimation \cite{Chesi}. 
Also, the reference \cite{GLMM} studied the same topic for a generic one-parameter model.
That is, the references \cite{Hall,Hassani,Chesi,GLMM}
did not studied the full unitary model $\SU(t)$.
The reference \cite{YRC} considered the first topic 
for the full unitary model $\SU(t)$ under a different context, i.e.,
the contest of gate programming.
However, these studies did not discuss the relation with 
the general framework of the representation theory
because their analyses rely on their specific models.
Moreover, they did not discuss the second topic.
In this paper, we study both topics 
under the multi-phase estimation model and 
the full unitary model $\SU(t)$.

To see the standard scaling for the mutual information, 
section \ref{S2} reviews existing result for state estimation.
Section \ref{S3} presents our result when a general unitary representation is given.
Section \ref{S4} applies the above result to the case with multi-phase estimation.
Section \ref{S5} applies the above result to the case with 
unitary group $\SU(t)$.
Section \ref{S6} gives the proof for a key statement in 
section \ref{S5}.
Section \ref{S7} makes conclusion.

\section{State estimation}\label{S2}
To see the standard scaling for the mutual information, 
we consider state estimation, i.e., 
we focus on a $d$-parameter state family 
$\{\rho_{\theta}\}_{\theta \in \Theta}$
on the Hilbert space ${\cal H}_A$, where
$\Theta$ is a subset of $\mathbb{R}^d$.
We consider a Bayesian prior $\mu$ on $\Theta$. 
We denote the set of density matrices on ${\cal H}$ by
${\cal S}({\cal H})$.
We denote the classical system $\Theta$ by $B$.
We consider $n$-copy state $\rho_\theta^{\otimes n}$. 
We have classical-quantum state
\begin{align}
\rho_{AB}:=\int_{\Theta} \rho_\theta^{\otimes n} \otimes |\theta\rangle \langle \theta|
\mu(d\theta). 
\end{align}
We focus on the mutual information between $A$ and $B$, which is given as
\begin{align}
I(A;B)&=
S(\rho_A)-\int_{\Theta} S(\rho_\theta^{\otimes n})\mu(d\theta),
\notag\\
&=\int_{\Theta} D(\rho_\theta^{\otimes n}\| \rho_A)
\mu(d\theta),
\end{align}
where $S(\rho)$ expresses the von Neumann entropy 
$-\Tr \rho \log \rho $ of $\rho$,
and $D(\rho\|\sigma)$ expresses the quantum relative entropy
$\Tr \rho(\log \rho-\log \sigma)$.

When all densities are commutative, i.e., the model is classical, the references 
\cite{KB1,KB2} showed that
\begin{align}
I(A;B)= \frac{d}{2}\log n+O(1).\label{NJR}
\end{align}

When our model $\{\rho_{\theta}\}_{\theta \in \Theta}$
is the full model on $t$-dimensional system
and $ \mu$ is invariant for unitary action, 
the reference \cite{H09} showed that 
\begin{align}
I(A;B)= \frac{t^2-1}{2}\log n+O(1).\label{NJR2}
\end{align}
Since the number of parameters of the full model is $t^2-1$,
\eqref{NJR2} can be considered as a generalization of 
\eqref{NJR}.
When a model satisfies a certain condition, 
using the result by \cite{YBCH},
we can show 
\begin{align}
I(A;B)= \frac{d}{2}\log n+o(\log n).\label{NJR3}
\end{align}
That is, the leading term is 
the number of parameters times 
$\frac{1}{2}\log n$.

We back to the spirit of population coding, and focus on the number 
$N(\{\rho^{\otimes n}_{\theta}\}_{\theta \in \Theta},\epsilon)$
of distinguishable states among $\{\rho^{\otimes n}_{\theta}\}_{\theta \in \Theta}$
with the average decoding error probability $\epsilon>0$.
As shown in \cite[(4.32)]{H2017QIT}, using Fano inequality, we can 
evaluate this number as
\begin{align}
\log N(\{\rho^{\otimes n}_{\theta}\}_{\theta \in \Theta},\epsilon)
\le \frac{\log 2 + I(A;B)}{1-\epsilon}.
\end{align}
That is, the relations \eqref{NJR}, \eqref{NJR2}, and \eqref{NJR3}
give upper bounds of this number.

In estimation theory, it is also important to clarify 
the distance between the estimate $\theta_{guess}$
and the true value $\theta_{true}$
when we exclude a certain error probability $\epsilon>0$;
\begin{align}
R_\epsilon:=\max_{R}\{R|
{\rm Pr}(\|\theta_{guess}-\theta_{true}\| > R)\le\epsilon\},
\label{SGI}
\end{align}
where $ \|x\|:= \max_j|x_j|$.
In the state estimation,
the normalized difference
$\sqrt{n}(\theta_{guess}-\theta_{true})$.
asymptotically obeys a Gaussian distribution,
which is called local asymptotic normality \cite{YBCH}.
This means that the above value $R_\epsilon$ scales $O(n^{-1/2})$ and its coefficient is asymptotically determined by the covariance of
$\sqrt{n}(\theta_{guess}-\theta_{true})$.
This relation is the reason why
we consider the minimum covariance 
in the parameter estimation in classical statistics.
Since this value scales as the square of variance,
we can consider that 
the value $R_\epsilon$ scales $O(n^{-1})$ in the Heisenberg scaling.
That is, when we focus on $R_\epsilon$,
the Heisenberg scaling can be considered as 
the case when the value $R_\epsilon$ scales $O(n^{-1})$.
However, under the Heisenberg scaling,
the variance $n(\theta_{guess}-\theta_{true})$
does not asymptotically obey a Gaussian distribution.
That is, its asymptotic distribution is not fixed.
Therefore, 
the coefficient of $R_\epsilon$ in the scale $O(n^{-1})$
cannot be asymptotically determined by the covariance of
${n}(\theta_{guess}-\theta_{true})$.
The minimization of the variance does not have the same important role under the Heisenberg scaling as under the usual scaling.

In the following, for simplicity, we consider the one-parameter case,
and assume that the parameter set is given as an interval $[a,b]$.
In this case, 
$\lfloor(b-a)/(2R_\epsilon)\rfloor$ elements of $[a,b]$
can be distinguished within the error probability $\epsilon$.
This relation between $R_\epsilon$ and the number of distinguished elements can be extended to the higher dimensional case.
That is, the number of distinguished elements
is meaningful in the parameter estimation
even under the Heisenberg scaling.

%hypercube

%Also, when a model certain conditions, using the result by \cite{YBCH}, we can show 
%\begin{align}
%I(A;B)= \frac{d}{2}\log n+o(\log n).\label{NJR3}
%\end{align}

\section{General problem formulation}\label{S3}
\subsection{General description}\label{S31}
Let $G$ be a compact group and $\mu$ be the Haar measure of $G$.
We focus on a unitary representation $f$ of $G$ over a 
finite-dimensional Hilbert space.
For this aim, we choose an input state.
We assume that any entangled input state with
any reference system is available.
To consider this problem, we denote the set of the labels of 
irreducible representations of $G$ by $\hat{G}$.
Let ${\cal U}_\lambda$ be the irreducible representation space 
identified by $\lambda \in \hat{G}$, and
$d_\lambda$ be its dimension.

We also define the twirling operation for the group $G$ as
\begin{align}
{\cal T}_G(\rho):=\int_{G}
 f(g')\rho f(g')^\dagger \mu(dg').
\end{align}
We denote the set of irreducible representations appearing in $f$
by $\hat{G}_f$.
We decompose the representation system ${\cal H}_A$ as follows.
\begin{align}
{\cal H}_A= \bigoplus_{\lambda \in \hat{G}_f}
{\cal U}_\lambda \otimes \mathbb{C}^{n_\lambda},
\end{align}
where 
$n_\lambda$ expresses the multiplicity of the representation 
space ${\cal U}_\lambda$.

When a reference system $\mathbb{C}^{l}$ is available, we have
\begin{align}
{\cal H}_{Al}:={\cal H}_A\otimes \mathbb{C}^{l}
= \bigoplus_{\lambda \in \hat{G}_f}
{\cal U}_\lambda \otimes \mathbb{C}^{l n_\lambda}.
\end{align}
However, when the input state is a pure state, 
the orbit is restricted to the following space by choosing 
a suitable subspace $\mathbb{C}^{\min(d_\lambda,l n_\lambda)}$
of $\mathbb{C}^{l n_\lambda}$.
That is, our representation space can be considered as follows.
\begin{align}
\bigoplus_{\lambda \in \hat{G}_f}
{\cal U}_\lambda \otimes \mathbb{C}^{\min(d_\lambda,l n_\lambda)}.
\end{align}
When $l \ge d_\lambda/n_\lambda$ for any $\lambda \in \hat{G}_f$,
our representation is given as
\begin{align}
{\cal H}_{AR}:=\bigoplus_{\lambda \in \hat{G}_f}
{\cal U}_\lambda \otimes \mathbb{C}^{d_\lambda}.\label{GH1}
\end{align}
In the following, we consider the above case.
We denote the projection to ${\cal U}_\lambda \otimes \mathbb{C}^{d_\lambda}$
by $P_\lambda$.

\subsection{Mutual information}\label{S32}
We choose an input state 
$\rho$
and a distribution $P$ on the system 
${\cal B}=G$, where 
the random choice of ${\cal B}$ is denoted by the random variable $B$.
So, we have a classical-quantum state 
$\sum_{g \in {\cal X}} P (g) 
f(g)\rho f(g)^\dagger \otimes |g\rangle\langle g| $.
%We also the final system ${\cal H}\otimes \mathbb{C}^l$ by $A_l$.
We focus on the mutual information $I(Al;B)$ or $(AR;B)$, which depends on $\rho$ and $P$.
When $\rho$ is decomposed as $\sum_j q_j \rho_j$,
the mutual information is evaluated as
\begin{align}
&I(Al;B)[P,  \rho]\notag\\
:=&
S\Big(\sum_{g' \in {\cal X}} P (g') 
f(g')\rho f(g')^\dagger \Big) \notag\\
&-
\sum_{g \in {\cal X}} P (g) 
S\Big(f(g)\rho f(g)^\dagger \Big)  \notag\\
=&
\sum_{g\in {\cal X}} P (g) \notag\\
&\cdot 
D\Big(f(g)\rho f(g)^\dagger \Big\|
\sum_{g' \in {\cal X}} P (g') 
f(g')\rho f(g')^\dagger \Big) \notag\\
\le &
\sum_j q_j
\sum_{g \in {\cal X}} P (g) \notag\\
&\cdot 
D\Big(f(g)\rho_j f(g)^\dagger \Big\|
\sum_{g' \in {\cal X}} P (g') 
f(g')\rho_j f(g')^\dagger \Big) .
\end{align}
The final inequality follows from the joint convexity of relative entropy.
Since any mixed state can be written as a mixture of pure states,
to maximize the mutual information, we can restrict the input state as an input pure state
$|\psi\rangle=
\oplus_{\lambda \in \hat{G}_f}\sqrt{p_\lambda} |\psi_\lambda\rangle$. 
In this case,
since the von Neumann entropy of a pure state is zero,
the mutual information is given as the von Neumann entropy of the mixture of the final states.
\begin{align}
&I(Al;B)[P,  |\psi\rangle]
\notag\\
=&
S\Big(\sum_{g' \in {\cal X}} P (g') 
f(g')|\psi\rangle \langle \psi|f(g')^\dagger \Big) .
\end{align}

We consider the following optimization problem.
\begin{align}
I(G)_{Al}:=\max_{|\psi\rangle} \max_{P} I(Al;B)[P,  |\psi\rangle] .
\end{align}
As shown in \cite{Korzekwa}, 
due to the convexity of von Neumann entropy 
the maximum is attained by the Haar measure $\mu$.
Hence, 
\begin{align}
\max_{P} I(Al;B)[P,  |\psi\rangle] =&
I(Al;B)[\mu,  |\psi\rangle] \notag\\
=&
S({\cal T}_G(  |\psi\rangle \langle \psi| )), \label{BC6}
\end{align}
which implies 
\begin{align}
I(G)_{Al} =\max_{|\psi\rangle} 
S({\cal T}_G(  |\psi\rangle \langle \psi| )).
\end{align}
Here, 
we set $|\phi_\lambda\rangle$ to be a pure state 
${\cal U}_\lambda \otimes 
\mathbb{C}^{\min(d_\lambda,l n_\lambda)}$ such that 
$\Tr_{{\cal U}_\lambda}|\phi_\lambda\rangle\langle \phi_\lambda|$
is the completely mixed state on 
$\mathbb{C}^{\min(d_\lambda,l n_\lambda)}$.
Such a state $|\phi_\lambda\rangle$ is called 
the maximally entangled state on 
${\cal U}_\lambda \otimes 
\mathbb{C}^{\min(d_\lambda,l n_\lambda)}$.

When $|\psi\rangle$ is the state 
$|\phi(p)\rangle:=\sum_{\lambda \in \hat{G}_f}
\sqrt{p_\lambda} |\phi_\lambda\rangle$, where
$p$ is a distribution on $\hat{G}_f$,
we have
\begin{align}
&S({\cal T}_G(  |\phi(p)\rangle \langle \phi(p)| )) \notag\\
=&S\Big( \sum_{\lambda \in \hat{G}_f}p_\lambda 
\frac{P_\lambda}{d_\lambda \min (n_\lambda, d_\lambda)}\Big)\notag\\ 
=&S(p)+\sum_{\lambda \in \hat{G}_f} p_\lambda 
\log (d_\lambda \min (n_\lambda, d_\lambda)).
\label{BC7}
\end{align}
Since the dimension of support of 
${\cal T}_G(  |\psi\rangle \langle \psi| )$ is upper bounded by
$\sum_{\lambda \in \hat{G}} 
d_\lambda \min (n_\lambda, d_\lambda)$,
we have
\begin{align}
&I(X;Al)[\mu,  |\psi\rangle] 
=  S({\cal T}_G(  |\psi\rangle \langle \psi| ) ) \notag\\
\le & R(G)_{Al}:=
\log  
\Big(\sum_{\lambda \in \hat{G}_f} 
d_\lambda \min (l n_\lambda, d_\lambda)
\Big) .
\end{align}
The equality holds
when the input state is 
$|\phi(p)\rangle$ and
$p_\lambda=
\frac{d_\lambda \min (l n_\lambda, d_\lambda)}
{\sum_{\lambda' \in \hat{G}_f} 
d_{\lambda'} \min (l n_{\lambda'}, d_{\lambda'})}$.
Therefore, we have
\begin{align}
I(G)_{Al}
= \log  
\Big(\sum_{\lambda \in \hat{G}_f} 
d_\lambda \min (l n_\lambda, d_\lambda)
\Big) .
\end{align}
Overall, the optimal input state is 
\begin{align}
\sum_{\lambda \in \hat{G}_f}
\sqrt{
\frac{d_\lambda \min (l n_\lambda, d_\lambda)}
{\sum_{\lambda' \in \hat{G}_f} 
d_{\lambda'} \min (l n_{\lambda'}, d_{\lambda'})}
} |\phi_\lambda\rangle.\label{BN1}
\end{align}
In particular, when $l$ satisfies the condition $l \ge d_\lambda/n_\lambda$ for any $\lambda \in \hat{G}_f$,
\begin{align}
I(G)_{AR}
=I(G)_{Al}
= R(G)_{AR}:=\log  
\Big(\sum_{\lambda \in \hat{G}_f} 
d_\lambda^2 
\Big) .
\end{align}

\subsection{Number of distinguishable elements}\label{S33}
Next, we consider how many elements in $G$ can be distinguished
via the above process.
When the input state is $\rho$, we denote the number of distinguishable states among $\{f(g) \rho f(g)^\dagger\}_{g \in G}$
with the average decoding error $\epsilon>0$ by $M_{Al}(\rho,\epsilon)$. 
That is,
\begin{align}
&M_{Al}(\rho,\epsilon)\notag\\
:=&
\max
\left\{M\middle| 
\begin{array}{l}
\frac{1}{M}
\sum_{j=1}^M 
\Tr f(g_j) \rho f(g_j)^{\dagger} \Pi_j \\
\ge 1-\epsilon
\end{array}
\right\},\Label{BNE}
\end{align}
where 
the maximum in \eqref{BNE} is taken with respect to 
${M,g_1,\ldots,g_M \in G,\{\Pi_j\}_{j=1}^M }$.
We consider the following optimization problems.
\begin{align}
M_\epsilon(G)_{Al}&:=\max_{\rho} 
M_{Al}(\rho ,\epsilon).
\end{align}

The number of distinguishable elements
is related to the error radius of the estimation.
We consider a distance $D$
distance on $G$ satisfying 
$D(g_1,g_2)=D(gg_1,gg_2)=D(g_1g,g_2g)$ for $g,g_1,g_2 \in G$.
As the unitary version of \eqref{SGI}, 
we focus on the following value.
\begin{align}
R_\epsilon:=\max_{R}\{R|
{\rm Pr}(D(g_{guess},g_{true}) > R)\le\epsilon\}.
\end{align}
To evaluate the above value, we introduce
\begin{align}
B(R):=\mu_H(\{g| D(g,g_{0}) \le  R\}).
\end{align}
Then, we have
\begin{align}
M_\epsilon(G)_{Al} B(R_\epsilon)
\le 1
\end{align}
That is, using $M_\epsilon(G)_{Al}$, we have the following upper bound for $R_\epsilon$;
\begin{align}
R_\epsilon \le B^{-1}( M_\epsilon(G)_{Al}^{-1}).\label{BNBB}
\end{align}

In addition, 
when $\{g_j\}$ are equally spaced, i.e., 
the subset $\{g_j\}$ forms a discrete subgroup of $G$,
the maximum value \eqref{BNE} is achieved with a certain $\epsilon>0$.
Given a discrete subgroup $\Omega:=\{g_j\} \subset G$,
we define 
$B(\Omega):=\mu_H (\{g\in G| \min_{g}D(g',g)=D(e,g) \})$.
When $|\Omega|=M_\epsilon(G)_{Al}$,
we have
an upper bound of $R_\epsilon$
\begin{align}
&R_\epsilon \le  R(\Omega)\notag\\
:=& \min_{g\in G}\{D(e,g)|\exists g'\in \Omega, 
D(e,g)=D(g',g)\}.
\end{align}

However, 
it is not so easy to find a relation between
a variance-type of error and $R_\epsilon$.
Indeed, in the case of state estimation,
the limiting distribution of the variable 
$\sqrt{n}(\theta_{guess}-\theta_{true})$ 
asymptotically obeys the Gaussian distribution,
the asymptotic variance derives the asymptotic behavior of $R_\epsilon$.
However, in the unitary estimation,
the asymptotic normality does not hold \cite{Imai_2009}.
Hence, we have no useful relation between
a variance-type of error and $R_\epsilon$.

To maximize the mutual information, we can restrict the input state as an input pure state $|\psi\rangle$
because a mixed input state can be simulated as a randomized choice of input pure states.
In this case, $M_{Al}(|\psi\rangle,\epsilon)$ can be evaluated by using 
the R\'{e}nyi entropy 
$S_\alpha\Big(\sum_{g' \in {\cal X}} P (g') 
f(g')|\psi\rangle \langle \psi|f(g')^\dagger \Big)$
of the mixture of the final states, where
$S_\alpha(\rho):=\frac{1}{\alpha-1}\log \Tr \rho^{\alpha}$ with $\alpha\neq 1$.
Using the results \cite[(6)]{Burnashev} and \cite[(4.67)]{H2017QIT}, we have
\begin{align}
&\log M(|\psi\rangle,\epsilon) \notag \\
\ge &
\max_{P}S_{\alpha}\Big(\sum_{g' \in {\cal X}} P (g') 
f(g')|\psi\rangle \langle \psi|f(g')^\dagger \Big)\notag \\
&-\frac{1}{\alpha-1}(\log 2 -\log \epsilon) 
\label{NM2} \\
&\log M(|\psi\rangle,\epsilon) \notag \\
\le &
\max_{P}S_{\beta}\Big(\sum_{g' \in {\cal X}} P (g') 
f(g')|\psi\rangle \langle \psi|f(g')^\dagger \Big)\notag \\
&+\frac{1}{\beta-1}\log (1-\epsilon) \label{NM1}
\end{align}
for $0< \beta<1$ and $1< \alpha\le 2$.
For its detailed derivation, see Appendix \ref{AP1}.
The above maximum is attained when $P$ is the Haar measure $\mu$.
Since the dimension of support of 
${\cal T}_G(  |\psi\rangle \langle \psi| )$ is upper bounded by
$\sum_{\lambda \in \hat{G}} 
d_\lambda \min (n_\lambda, d_\lambda)$,
we have
\begin{align}
S_\alpha({\cal T}_G(  |\psi\rangle \langle \psi| ) )
\le R(G)_{Al}.
\end{align}
The equality holds
when the input state is $|\phi\rangle$.
Thus, choosing the best choice $\alpha=2$ and taking the limit
$\beta\to  +0$, we have
\begin{align}
& R(G)_{Al}
 -(\log 2 -\log \epsilon)
\le \log M_\epsilon(G)_{Al} \notag \\
\le &
R(G)_{Al} - \log (1-\epsilon) .
\end{align}
In particular, we have
\begin{align}
&R(G)_{AR}
 -(\log 2 -\log \epsilon)\notag \\
\le &\log M_\epsilon(G)_{AR} \notag \\
\le &\log  R(G)_{AR}
-\log (1-\epsilon) .\label{VB3}
\end{align}
Therefore, when $1>\epsilon>0$ is fixed,
$\log  R(G)_{AR}$ is the dominant term.

\section{Multiple phase estimation}\label{S4}
Here, we consider the multi-phase application model on 
the $t$-dimensional system ${\cal H}_A$ spanned by
$\{|j\rangle\}_{j=0}^{t-1} $.
This model is given as the application of 
the group 
$MP_{t-1}:=\{U_\theta\}_{\theta\in [0,2\pi)^{t-1}}$, where the unitary 
$U_\theta$ is defined as 
$|0\rangle \langle 0|
+\sum_{j=1}^{t-1}
e^{i \theta_j}|j\rangle \langle j|$.
Now, we consider the $n$-parallel application of  
$U_\theta$ on ${\cal H}_{A^n}:={\cal H}_A^{\otimes n}$.
In this representation, 
all irreducible representations are 
characterized by 
the combinatorics ${\cal C}^n_t$, which is defined as
\begin{align}
{\cal C}^n_t:= \Big\{\vec{n}:=(n_0,n_1,\ldots, n_{t-1})\in \mathbb{N}^t \Big|
\sum_{j=0}^{t-1} n_j=n\Big\},
\end{align}
where $ \mathbb{N}$ is the set of non-negative integers.
Since all irreducible representations are one-dimensional, 
$d_{\vec{n}}=1$ for $\vec{n} \in {\cal C}^n_t$ so that
we do not need to attach any reference system, i.e., 
$R(G)_{A1}=R(G)_{AR}$.
Since $|{\cal C}^n_t|={n+t-1\choose t-1}$,
we have
\begin{align}
I(MP_{t-1})_{A^n}=&
R( MP_{t-1})_{A^n}
=\log {n+t-1\choose t-1}\notag \\
=& (t-1)\log n+o(1)\label{NM4}
\end{align}
while the number of our parameters is $t-1$. 
Compare with \eqref{NJR3}, 
\eqref{NM4} achieves the twice of the state estimation case.
In addition, due to \eqref{VB3}, the value \eqref{NM4} expresses 
the logarithm of the number of
distinguishable elements in $MP_{t-1}$.
Hence, we can say that 
multi-phase estimation achieves Heisenberg scaling 
in both senses, the mutual information and 
the logarithm of the number of distinguishable elements.

Under this model, the references \cite{Hall,Hassani,Chesi} derived 
Heisenberg scaling for the mutual information,
but it did not study 
the number of
distinguishable elements, which is essential for population coding.

When $t=2$, the optimal input state is 
\begin{align}
\sum_{j=0}^n \frac{1}{\sqrt{n+1}} |\overbrace{0\ldots 0}^{j}\overbrace{1\ldots 1}^{n-j}\rangle. \label{BS4}
\end{align}
The asymptotic estimation error with the above input state has been studied in
\cite[Sections 3 and 4]{Imai_2009}.
As explained in \cite[Section 4]{HLY}, this method is essentially the same as Kitaev's method \cite{Kitaev,Cleve},
and has been implemented by \cite{Higgins2007}.

Here, we compare the input state \eqref{BS4} with the case with the noon state \cite{Giovannetti,Giovannetti06,Giovannetti11}.
When the input state is the noon state,
we employ only two irreducible representations to maximize the Fisher information. That is, the input state belongs to the subspace spanned by 
$|0\rangle^{\otimes n}$ and $|1\rangle^{\otimes n}$.
The relations \eqref{BC6} and \eqref{BC7} imply
\begin{align}
I(X;A1)[\mu,  
\frac{1}{\sqrt{2}}(|0\rangle^{\otimes n}+|1\rangle^{\otimes n}
)]=\log 2. 
\end{align}
This equation shows that
the input noon state has quite small global information,
which certifies the impossibility of global estimation by 
the input noon state.

However, 
when we focus on the error $\sin^2\frac{\theta_{guess}-\theta_{true}}{2}$,
the input state \eqref{BS4} does not coincide with the optimal input
\cite[Sections 3 and 4]{Imai_2009}, \cite[Section 10.2]{Fourier}.
Further, the input state \eqref{BS4} does not achieve
Heisenberg limit in this sense \cite[Sections 3 and 4]{Imai_2009}, \cite{BHBMPW}.
However, the references \cite[Sections 3 and 4]{Imai_2009} 
\cite[Section 4]{HLY}
show that the input state \eqref{BS4} achieves the Heisenberg scaling  
in the sense limiting distribution.

\section{$n$-tensor product representation of $\SU(t)$}\label{S5}
We consider the case when the representation is given as
$n$-tensor product of $\SU(t)$
on ${\cal H}_A=\mathbb{C}^t$.
By using the set $Y_t^n$ of Semistandard Young tableau, 
our representation is given as 
\begin{align}
{\cal H}_{A^n}=\bigoplus_{\lambda \in Y_t^n} 
{\cal H}_{\lambda} \otimes {\cal M}_\lambda,
\end{align}
where ${\cal M}_\lambda= \mathbb{C}^{n_\lambda}$
expresses the multiplicity space of the irreducible unitary 
presentation $U^\lambda$,
and the integer $n_\lambda$ is the multiplicity.

When $\lambda=(\lambda_1, \ldots, \lambda_{t})$
with $\lambda_1\le \ldots\le \lambda_{t} $ and $
\sum_j \lambda_j=n$,
$d_\lambda$ is given as \cite[(4.46)]{H-group2}
\begin{align}
d_\lambda=\prod_{1 \le j<k\le t }\frac{k-j+\lambda_k-\lambda_j}{k-j}.
\end{align}
It is upper bounded by
$(n+1)^{t(t-1)/2}$.
When the reference system $\mathbb{C}^l$
is sufficiently large in the sense of \eqref{GH1},
the maximum mutual information is evaluated as
\begin{align}
&I(\SU(t))_{A^nR}=\!
R( \SU(t))_{A^nR}=\!
\log \Big(\sum_{\lambda' \in Y_{t}^n }d_{\lambda'}^2 \Big)
\notag \\
\le &
\log (n+1) ^{(t+1)t+ (t-1)}=
(t^2-1)\log (n+1)
\label{M24}
\end{align}
because $|Y_{t}^n|\le (n+1) ^{t-1}$.

Due to \eqref{BN1}, the optimal input state is 
\begin{align}
\sum_{\lambda \in Y_{t}^n}
d_\lambda
\sqrt{
\Big(\sum_{\lambda' \in Y_{t}^n} 
d_{\lambda'}^2\Big)^{-1}} |\phi_\lambda\rangle\label{BN2},
\end{align}
Here, $|\phi_\lambda\rangle$ is the maximally entangled state on 
${\cal H}_{\lambda} \otimes {\cal M}_\lambda$.

When $t$ is fixed and only $n$ increases, 
we have
\begin{align}
\log \Big(\sum_{\lambda' \in Y_{t}^n }d_{\lambda'}^2 \Big)
=(t^2-1)\log n+O(1),\label{Hei}
\end{align}
which is shown in Section \ref{S6}.
Since $t^2-1$ is the number of parameters of $\SU(t)$.
The leading term in \eqref{Hei} is 
twice of the case of state estimation case given in \eqref{NJR3}.
Therefore, this behavior can be considered to achieve the Heisenberg scaling in the sense of mutual information.

At least, when $t=2$, 
we can show the above relation as follows.
Consider the case with $t=2$.
When $n=2m$,
we have
\begin{align}
&\sum_{\lambda' \in Y_{d}^n }d_{\lambda'}^2 
= \sum_{k=0}^m(2k+1)^2\notag \\
=& \frac{(m+1)(2m+1)(2m+3)}{3}.
\end{align}
When $n=2m-1$,
we have
\begin{align}
\sum_{\lambda' \in Y_{d}^n }d_{\lambda'}^2 
= \sum_{k=1}^m (2k)^2
= \frac{2m(m+1)(2m+1)}{3}.
\end{align}
Then, we have
\begin{align}
\log \Big(\sum_{\lambda' \in Y_{d}^n }d_{\lambda'}^2 \Big)
= 3 \log n -\log 6 +O(\frac{1}{n}).\label{BN9}
\end{align}

Here, we compare the above optimal case \eqref{M24},\eqref{Hei}
with the case of the input state maximizing the trace of 
Fisher information matrix \cite{Imai}.
The maximally entangled state $|\Phi\rangle$
on the symmetric subspace 
maximizes the trace of Fisher information matrix \cite{Imai}.
In this case, 
since the dimension of the symmetric subspace is 
${n+t-1\choose t-1}$,
the relations \eqref{BC6} and \eqref{BC7} imply
\begin{align}
&I(X;AR)[\mu,  |\Phi\rangle]=2 \log {n+t-1\choose t-1}\notag\\
=&2(t-1)\log n+o(1),
\end{align}
which is much smaller than \eqref{M24},\eqref{Hei}.

Next, we focus on $M_\epsilon(\SU(t))_{AR}$.
Due to \eqref{VB3}, \eqref{M24}, and \eqref{Hei},
$\log M_\epsilon(\SU(t))_{AR}$ behaves as
\begin{align}
\log M_\epsilon(\SU(t))_{AR}
=(t^2-1)\log n+O(1).
\end{align}
Since $\SU(t)$ is parametrized by $t^2-1$ parameters,
$B(R)$ scales as $O( R^{t^2-1})$. 
Since $M_\epsilon(\SU(t))_{AR}$ scales as $O(n^{t^2-1})$,
the upper bound of $R_\epsilon$ scales as $O(n^{-1})$.

\section{Proof of (\ref{Hei})}\Label{S6}
Since we have \eqref{M24}, it is sufficient to show
\begin{align}
\log \Big(\sum_{\lambda' \in Y_{t}^n }d_{\lambda'}^2 \Big)
\ge
(t^2-1)\log n+O(1)\label{NBE}.
\end{align}

We choose a positive real number $a_n$ such that $\frac{n}{a_n t}$ is an integer.
We define the subset 
$Y_t^n(a_n)\subset Y_t^n$ as
\begin{align}
&Y_t^n(a_n)\notag\\
:=&
\Big\{\lambda
\Big| \frac{n}{a_n t} \le \lambda_{k+1}-\lambda_k 
\hbox{ for } k=1, \ldots, t-1 \Big\}.
\end{align}
For $\lambda \in Y_t^n(a_n)$, we have
\begin{align}
d_\lambda
=&
\prod_{1 \le j<k\le t }\frac{k-j+\lambda_k-\lambda_j}{k-j}\notag\\
\ge &
\frac{1}{t}\prod_{1 \le j<k\le t } \frac{\lambda_k-\lambda_j}{t} \notag\\
\ge &
\prod_{1 \le j<k\le t } \frac{n}{a_n t^2}
=
\big(\frac{n}{a_n t^2}\big)^{\frac{(t-1)t}{2}}.
\end{align}

Also, using $m_n:= n- \frac{n}{a_n}$,
we have
\begin{align}
|Y_t^n(a_n)|= {m_n+t-1 \choose t-1}
\ge \frac{m_n^{t-1}}{(t-1)\!}.
\end{align}
Therefore,
\begin{align}
&\sum_{\lambda \in Y_{t}^n }d_{\lambda}^2
\ge \!
\sum_{\lambda \in Y_{t}^n(a_n) }d_{\lambda}^2\!
\ge \!
\sum_{\lambda \in Y_{t}^n(a_n) }
\big(\frac{n}{a_n t^2}\big)^{(t-1)t} \notag\\
=&|Y_t^n(a_n)|
\big(\frac{n}{a_n t^2}\big)^{(t-1)t} 
\ge \frac{m_n^{t-1}}{(t-1)\!}
\big(\frac{n}{a_n t^2}\big)^{(t-1)t} .
\end{align}
When $a_n$ choose as a value such that $2 \le a_n \le 3$,
we have
\begin{align}
&\log 
\Big(\frac{m_n^{t-1}}{(t-1)\!}
(\frac{n}{a_n t^2})^{(t-1)t} \Big)
=(t^2-1)\log n+O(1)
\end{align}
because $t-1+ (t-1)t=t^2-1$.
Hence, we obtain \eqref{NBE}.

\section{Conclusion}\label{S7}
We have derived general formulas for 
the mutual information 
and the logarithm of the number of distinguishable elements 
when a unitary group representation is given.
We have proposed the above quantities 
as a figure of merit to address the population coding with group representation
because these quantities reflect the global information structure unlike Fisher information.
Then, we have applied these general formulas
to the case with multi-phase estimation and multiple applications of $\SU(t)$.
As the results, we have revealed that 
the optimal strategy realized the twice value of the 
standard case for these two quantities, which can be considered as Heisenberg scaling.
We have also shown that
the optimal strategies for maximizing Fisher information
have much smaller values for these quantities.
This fact shows the advantage of our figure of merit over Fisher information.

\section*{Acknowledgement}
The author is supported in part by the National Natural Science Foundation of China (Grant No.
62171212).

\appendix
\section{Derivations of \eqref{NM2} and \eqref{NM1}}\label{AP1}
Since $f(g')|\psi\rangle \langle \psi|f(g')^\dagger$ is a pure state,
the reference \cite[(6)]{Burnashev} 
shows the inequality
\begin{align}
\epsilon
\le & 2(M(|\psi\rangle,\epsilon)-1)^s \notag\\
&\cdot\Tr \Big(\sum_{g' \in {\cal X}} P (g') 
f(g')|\psi\rangle \langle \psi|f(g')^\dagger \Big)^{1+s}\notag \\
\le & 2 M(|\psi\rangle,\epsilon)^s \notag\\
&\cdot\Tr \Big(\sum_{g' \in {\cal X}} P (g') 
f(g')|\psi\rangle \langle \psi|f(g')^\dagger \Big)^{1+s} 
\end{align}
for $0\le s\le 1$ and any distribution $P$ on $G$.
Choosing $\alpha=1+s$, we have 
\begin{align}
\log M(|\psi\rangle,\epsilon) %\notag \\
\ge &
S_{\alpha}\Big(\sum_{g' \in {\cal X}} P (g') 
f(g')|\psi\rangle \langle \psi|f(g')^\dagger \Big)\notag \\
&-\frac{1}{\alpha-1}(\log 2 -\log \epsilon) 
\end{align}
for $1< \alpha \le 2$.
Taking the maximum for $P$, we obtain \eqref{NM2}.

Since $f(g')|\psi\rangle \langle \psi|f(g')^\dagger$ is a pure state,
the reference \cite[(4.67)]{H2017QIT}
shows the inequality
\begin{align}
&\log (1-\epsilon) \notag\\
\le&
\max_{P}\log \Tr 
\Big(\sum_{g' \in {\cal X}} P (g') 
f(g')|\psi\rangle \langle \psi|f(g')^\dagger \Big)^{\frac{1}{1-s}}
\notag \\
&+\frac{s}{1-s}
\log M(|\psi\rangle,\epsilon) 
\end{align}
for $s<0$.
Choosing $\beta=\frac{1}{1-s}$, we have 
\eqref{NM1}.

\bibliographystyle{quantum}
\bibliography{references}
\end{document}